\documentclass[aps,prd,twocolumn,superscriptaddress,showpacs]{revtex4}
\usepackage[dvips]{graphicx}
\usepackage{amsmath,amssymb,times}

\newcommand{\bequ}{\begin{equation}}
\newcommand{\eequ}{\end{equation}}
\newcommand{\bea}{\begin{eqnarray}}
\newcommand{\eea}{\end{eqnarray}}

\DeclareSymbolFont{boldletters}{OML}{cmm} {b}{it}
\DeclareSymbolFontAlphabet{\mathbit}{boldletters}
\DeclareMathSymbol{\alpha}{\mathalpha}{letters}{"0B}
\DeclareMathSymbol{\beta}{\mathalpha}{letters}{"0C}
\DeclareMathSymbol{\gamma}{\mathalpha}{letters}{"0D}
\DeclareMathSymbol{\delta}{\mathalpha}{letters}{"0E}
\DeclareMathSymbol{\epsilon}{\mathalpha}{letters}{"0F}
\DeclareMathSymbol{\zeta}{\mathalpha}{letters}{"10}
\DeclareMathSymbol{\eta}{\mathalpha}{letters}{"11}
\DeclareMathSymbol{\theta}{\mathalpha}{letters}{"12}
\DeclareMathSymbol{\iota}{\mathalpha}{letters}{"13}
\DeclareMathSymbol{\kappa}{\mathalpha}{letters}{"14}
\DeclareMathSymbol{\lambda}{\mathalpha}{letters}{"15}
\DeclareMathSymbol{\mu}{\mathalpha}{letters}{"16}
\DeclareMathSymbol{\nu}{\mathalpha}{letters}{"17}
\DeclareMathSymbol{\xi}{\mathalpha}{letters}{"18}
\DeclareMathSymbol{\pi}{\mathalpha}{letters}{"19}
\DeclareMathSymbol{\rho}{\mathalpha}{letters}{"1A}
\DeclareMathSymbol{\sigma}{\mathalpha}{letters}{"1B}
\DeclareMathSymbol{\tau}{\mathalpha}{letters}{"1C}
\DeclareMathSymbol{\upsilon}{\mathalpha}{letters}{"1D}
\DeclareMathSymbol{\phi}{\mathalpha}{letters}{"1E}
\DeclareMathSymbol{\chi}{\mathalpha}{letters}{"1F}
\DeclareMathSymbol{\psi}{\mathalpha}{letters}{"20}
\DeclareMathSymbol{\omega}{\mathalpha}{letters}{"21}
\DeclareMathSymbol{\varepsilon}{\mathalpha}{letters}{"22}
\DeclareMathSymbol{\vartheta}{\mathalpha}{letters}{"23}
\DeclareMathSymbol{\varpi}{\mathalpha}{letters}{"24}
\DeclareMathSymbol{\varrho}{\mathalpha}{letters}{"25}
\DeclareMathSymbol{\varsigma}{\mathalpha}{letters}{"26}
\DeclareMathSymbol{\varphi}{\mathalpha}{letters}{"27}
\DeclareMathSymbol{\Gamma}{\mathalpha}{letters}{"00}
\DeclareMathSymbol{\Delta}{\mathalpha}{letters}{"01}
\DeclareMathSymbol{\Theta}{\mathalpha}{letters}{"02}
\DeclareMathSymbol{\Lambda}{\mathalpha}{letters}{"03}
\DeclareMathSymbol{\Xi}{\mathalpha}{letters}{"04}
\DeclareMathSymbol{\Pi}{\mathalpha}{letters}{"05}
\DeclareMathSymbol{\Sigma}{\mathalpha}{letters}{"06}
\DeclareMathSymbol{\Upsilon}{\mathalpha}{letters}{"07}
\DeclareMathSymbol{\Phi}{\mathalpha}{letters}{"08}
\DeclareMathSymbol{\Psi}{\mathalpha}{letters}{"09}
\DeclareMathSymbol{\Omega}{\mathalpha}{letters}{"0A}

\begin{document}
\preprint{SAGA-HE-257}
\title{Entanglement between deconfinement transition and 
chiral symmetry restoration 
}

\author{Yuji Sakai}
\email[]{sakai@phys.kyushu-u.ac.jp}
\affiliation{Department of Physics, Graduate School of Sciences, Kyushu University,
             Fukuoka 812-8581, Japan}

\author{Takahiro Sasaki}
\email[]{sasaki@phys.kyushu-u.ac.jp}
\affiliation{Department of Physics, Graduate School of Sciences, Kyushu University,
             Fukuoka 812-8581, Japan}

\author{Hiroaki Kouno}
\email[]{kounoh@cc.saga-u.ac.jp}
\affiliation{Department of Physics, Saga University,
             Saga 840-8502, Japan}

\author{Masanobu Yahiro}
\email[]{yahiro@phys.kyushu-u.ac.jp}
\affiliation{Department of Physics, Graduate School of Sciences, Kyushu University,
             Fukuoka 812-8581, Japan}

\date{\today}

\begin{abstract}
We extend the Polyakov-loop extended Nambu--Jona-Lasinio (PNJL) model by 
introducing an effective four-quark vertex depending on Polyakov loop. 
The effective vertex generates entanglement interactions between 
Polyakov loop and chiral condensate. 
The new model is consistent with lattice QCD data at 
imaginary quark-number chemical potential and real and imaginary isospin 
chemical potentials, particularly on strong correlation between 
the chiral and deconfinement transitions and also on 
the quark-mass dependence of the order of the Roberge-Weiss endpoint. 
We investigate the influence of the entanglement interactions 
on the location of the tricritical point at real isospin chemical potential 
and on the location of the critical endpoint at real quark-number chemical 
potential. 
\end{abstract}

\maketitle

\section{Introduction}
\label{Introduction}

An important query on the thermodynamics of quantum chromodynamics (QCD) 
is whether the chiral-symmetry restoration and 
the confinement-to-deconfinement transition take place simultaneously or not. 
If the two transitions do not coincide, phases such as 
the constituent quark phase~\cite{Cleymans,Kouno1} or 
the quarkyonic phase~\cite{McLerran1,Hidaka} may appear.

If the chiral and deconfinement transitions are of first order, 
discontinuities appear 
simultaneously in their (approximate) order parameters, that is, 
the chiral condensate $\sigma$ and the Polyakov loop $\Phi$
~\cite{BCGG,Kashiwa5}. 
Furthermore, if a nontrivial critical endpoint (CEP) exists 
at finite temperature ($T$) and quark-number chemical potential 
$\mu_{\rm q}$~\cite{AY}, 
susceptibilities of $\sigma$, $\Phi$ and other quantities diverge 
simultaneously~\cite{Fujii}. This indicates a coincidence of 
second-order phase transitions. 
At zero $\mu_{\rm q}$, the chiral and deconfinement transitions are 
found to be crossover \cite{KL1994}. Hence, 
there is no a priori reason why the two transitions coincide exactly. 
Actually, in lattice QCD (LQCD) simulations at zero chemical 
potential~\cite{KL1994,SAoki1998,YAoki2006}, 
there is a debate as to whether the transitions really coincide or not; 
see Ref.~\cite{Borsanyi} and references therein. 
LQCD simulations are far from perfection 
at real $\mu_{\rm q}$ because of the well-known sign problem~\cite{Kogut}. 
Fortunately, LQCD data are 
available at imaginary $\mu_{\rm q}$~\cite{FP,FP3,Elia,Chen34,Chen,D'Elia-3,FP2010,Nagata,Takaishi} and real and imaginary isospin chemical potential 
$\mu_{\rm I}$~\cite{Kogut2,Cea,D'Elia-2}, 
since LQCD has no sign problem there. 
The data show 
that chiral and deconfinement crossover transitions 
coincide within the numerical accuracy. 
Since there is no general reason 
for exact coincidence between crossover transitions, 
it is natural to think that the chiral and deconfinement crossover transitions 
nearly coincide as a result of strong correlation (entanglement) 
between $\sigma$ and $\Phi$. 
We investigate this possibility in the present paper. 

As an approach complementary to first-principle LQCD, 
we can consider effective models such as the  Nambu--Jona-Lasinio (NJL) 
model~\cite{NJ1,KHKB,AY,Fujii,KKKN,Osipov,Kashiwa} and 
the Polyakov-loop extended Nambu--Jona-Lasinio (PNJL) 
model~\cite{Meisinger,Dumitru,Fukushima-1,Fukushima-2,Ratti,Ghos,Megias,Ciminale,Rossner,Hansen,Sasaki,Schaefer,Costa,Kashiwa1,Fu,Abuki,Sakai,Sakai1,Sakai2,Kashiwa5,Kouno,Sakai3,Matsumoto,Sasaki-T,Sakai4}. 
The NJL model can describe chiral symmetry breaking, 
but not the confinement mechanism. 
The PNJL model is designed \cite{Fukushima-1} to make it possible 
to treat both the mechanisms. 
The PNJL model can reproduce results of LQCD at zero and imaginary $\mu _{\rm q}$~\cite{Sakai,Sakai1,Sakai2,Kouno}
 where LQCD has no sign problem. \\
At imaginary $\mu_{\rm q}=i\theta_{\rm q} T$, 
the grand canonical partition function 
$Z_{\rm GC}(\theta_{\rm q} )$ of QCD 
is related to the thermodynamic potential $\Omega_{\rm QCD}$ as 
$\Omega_{\rm QCD}(\theta_{\rm q})=-T \ln(Z_{\rm GC}(\theta_{\rm q}))$, 
where $\theta_{\rm q}$ is a real parameter. 
Roberge and Weiss (RW) found~\cite{RW} that QCD has a periodicity 
$\Omega_{\rm QCD}(\theta_{\rm q})=\Omega_{\rm QCD}(\theta_{\rm q} +2\pi k/3)$, 
showing that $\Omega_{\rm QCD}(\theta_{\rm q} +2\pi k/3)$ is transformed into 
$\Omega_{\rm QCD}(\theta_{\rm q})$ by the ${\mathbb Z}_3$ transformation with 
integer $k$. 
This means that QCD is invariant under a combination of the ${\mathbb Z}_3$ transformation and a parameter transformation $\theta_{\rm q} \to \theta_{\rm q}+2k\pi/3$~\cite{Sakai,Sakai1}, 
\bea
q &\to& Uq, \quad 
A_{\nu} \to UA_{\nu}U^{-1} - i/g (\partial_{\nu}U)U^{-1} , \nonumber \\ 
\theta _{\rm q}&\to& \theta _{\rm q}+2\pi k/3 ,
\label{ez3}
\eea
where $U(x,\tau)$ are elements of SU(3) with 
$
U(x,\beta=1/T)=\exp(-2i \pi k/3)U(x,0) ,
$
$q$ is the quark field and $A_{\nu}$ is the gauge field. 
We call this combination extended ${\mathbb Z}_3$ transformation. 
Thus, $\Omega_{\rm QCD}(\theta _{\rm q})$ has the extended ${\mathbb Z}_3$ symmetry, 
and hence quantities invariant under the extended ${\mathbb Z}_3$ 
transformation have RW periodicity~\cite{Sakai,Sakai1}. 
At the present stage, the PNJL model is only a realistic effective model 
that possesses both extended ${\mathbb Z}_3$ symmetry and 
chiral symmetry~\cite{Sakai,Sakai1}. 
Furthermore, the PNJL model can reproduce the 
first-order RW 
transition~\cite{RW} that occurs at $\theta _{\rm q}=(2k+1)\pi /3$ when 
$T$ is larger than some critical temperature $T_{\rm E}$. 
This property makes it possible to compare PNJL results with LQCD data 
quantitatively at imaginary $\mu_{\rm q}$. 
A current topic at imaginary $\mu_{\rm q}$ is 
what the order of the RW transition is at the endpoint $T=T_{\rm E}$. 
The recent LQCD simulations show that 
it is first-order for small and larger quark masses, but 
the order is weakened and could be second order 
at intermediate masses~\cite{D'Elia-3,FP2010}.

In the PNJL model, the correlation between $\sigma$ and $\Phi$ is weak, 
so that the chiral and deconfinement crossover transitions do not coincide 
without any fine-tuning of parameters~\cite{Sakai2}. 
For zero chemical potential, 
the scalar-type eight-quark interaction is necessary 
to obtain a coincidence between the two transitions, and 
for imaginary $\mu_{\rm q}$ the vector-type four-quark 
interaction is needed~\cite{Sakai2}. 
This fact indicates that a true correlation between $\sigma$ and $\Phi$ 
is stronger than that in the PNJL model appearing 
through the covariant derivative 
between quark and gauge fields. 
Actually, recent analyses~\cite{Braun,Kondo} 
based on the exact renormalization-group (ERG) equation~\cite{Wetterich} 
indicate that entanglement interactions between $\sigma$ and $\Phi$ 
appear in 
addition to the original entanglement through the covariant derivative.

In this paper, we extend the PNJL model by introducing 
an effective four-quark vertex depending phenomenologically on $\Phi$. 
The effective vertex generates entanglement interactions 
between $\sigma$ and $\Phi$. 
The functional form of the entanglement vertex is determined by 
respecting the extended ${\mathbb Z}_3$ symmetry and the chiral symmetry. 
The strength of the vertex is 
determined from LQCD data at imaginary $\mu_{\rm q}$, and 
the validity of the model setting is confirmed 
for real and imaginary values of $\mu_{\rm I}$ 
by comparing the model results with LQCD data. 
The new model is consistent with all LQCD data 
at imaginary $\mu_{\rm q}$ and real and imaginary $\mu_{\rm I}$. 
Particularly, the new model can reproduce 
two phenomena simultaneously; one is the strong correlation between 
the deconfinement and chiral transitions and the other is 
the quark-mass dependence of the RW endpoint 
predicted by LQCD very recently~\cite{D'Elia-3,FP2010}. 
We also analyze the influence of the entanglement interactions 
on the location of the tricritical point (TCP) in 
$\mu_{\rm I}$-$T$ plane and the location of the critical endpoint (CEP) in 
$\mu_{\rm q}$-$T$ plane. 
The present phenomenological approach is complementary to the ERG approach 
mentioned above.

In Sec. II, we explain the PNJL model briefly and introduce 
an effective four-quark vertex depending on $\Phi$. 
In Sec. III, the new model with the effective vertex is applied 
to the imaginary $\mu_{\rm q}$ region 
and the real and imaginary $\mu_{\rm I}$ regions and compared with LQCD there. 
Sec. IV is devoted to a summary. 

\section{PNJL model}
\label{PNJL}

We start with the standard two-flavor PNJL Lagrangian~\cite{Fukushima-1,Ratti}
\begin{align}
 {\cal L}  =& {\bar q}(i \gamma_\nu D^\nu -m_0)q \notag\\
             &\hspace{3mm} + G_{\rm s}[({\bar q}q)^2 
                          +({\bar q}i\gamma_5 {\vec \tau}q)^2] 
              - {\cal U}(\Phi [A],{\Phi} [A]^*,T) ,
             \label{eq:E1}
\end{align}
where $q$ denotes the two-flavor quark field, 
$m_0$ denotes the current quark mass, 
and $D^\nu=\partial^\nu+iA^\nu-i\mu_{\rm q}\delta^{\nu}_{0}$. 
Field $A^\nu$ is defined as $A^\nu=\delta^{\nu}_{0}gA^0_a{\lambda^a\over{2}}$, 
with gauge fields $A^\nu_a$, the Gell-Mann matrix $\lambda_a$, and the 
gauge coupling $g$.
In the NJL sector, $G_{\rm s}$ denotes the coupling constant of 
the scalar-type four-quark interaction. 
The Polyakov potential ${\cal U}$, defined in (\ref{eq:E13}), is a function 
of Polyakov loop $\Phi$ and its Hermitian conjugate $\Phi^*$,
\begin{align}
\Phi      = {1\over{N_{\rm c}}}{\rm Tr} L,~~~~
\Phi^{*}  = {1\over{N_{\rm c}}} {\rm Tr}L^\dag ,
\end{align}
with
\begin{align}
L({\bf x}) = {\cal P} \exp\Bigl[
                {i\int^\beta_0 d \tau A_4({\bf x},\tau)}\Bigr],
\end{align}
where ${\cal P}$ is the path ordering and $A_4 = iA_0 $. 
In the chiral limit ($m_0=0$), the Lagrangian density has the exact $SU(N_f)_{\rm L} \times SU(N_f)_{\rm R}\times U(1)_{\rm v} \times SU(3)_{\rm c}$  symmetry. 
The temporal component of the gauge field is diagonal in flavor space, 
because color and flavor spaces are completely separated 
in the present case. 
In the Polyakov gauge, $L$ can be written in a diagonal form 
in color space~\cite{Fukushima-1}: 
\begin{align}
L 
=  e^{i \beta (\phi_3 \lambda_3 + \phi_8 \lambda_8)}
= {\rm diag} (e^{i \beta \phi_a},e^{i \beta \phi_b},
e^{i \beta \phi_c} ),
\label{eq:E6}
\end{align}
where $\phi_a=\phi_3+\phi_8/\sqrt{3}$, $\phi_b=-\phi_3+\phi_8/\sqrt{3}$
and $\phi_c=-(\phi_a+\phi_b)=-2\phi_8/\sqrt{3}$. 
The Polyakov loop $\Phi$ is an exact order parameter of spontaneous 
${\mathbb Z}_3$ symmetry breaking in pure gauge theory.
Although ${\mathbb Z}_3$ symmetry is not an exact one 
in the system with dynamical quarks, it still seems to be a good indicator of 
the deconfinement phase transition. 
Therefore, we use $\Phi$ to define the deconfinement phase transition.

Making the mean field approximation and performing 
the path integral over the quark field, 
one can obtain the thermodynamic potential $\Omega$ (per volume), 
\begin{align}
\Omega =& -2 N_f \int \frac{d^3{\rm p}}{(2\pi)^3}
         \Bigl[ 3 E ({\bf p}) \nonumber\\
        & + \frac{1}{\beta}
           \ln~ [1 + 3(\Phi+\Phi^{*} e^{-\beta E^-({\bf p})}) 
           e^{-\beta E^-({\bf p})}+ e^{-3\beta E^-({\bf p})}]
         \nonumber\\
        & + \frac{1}{\beta} 
           \ln~ [1 + 3(\Phi^{*}+{\Phi e^{-\beta E^+({\bf p})}}) 
              e^{-\beta E^+({\bf p})}+ e^{-3\beta E^+({\bf p})}]
	      \Bigl]\nonumber\\
        & +U_{\rm M}+{\cal U}, 
\label{eq:E12} 
\end{align}
where $\sigma = \langle \bar{q}q \rangle $, 
$\Sigma_{\rm s} = -2 G_{\rm s} \sigma$, $M=m_0 + \Sigma_{\rm s}$, 
$U_{\rm M}= G_{\rm s} \sigma^2$, $E({\bf p})=\sqrt{{\bf p}^2+M^2}$ and 
$E^\pm({\bf p})=E({\bf p})\pm \mu_{\rm q} =E({\bf p})\pm i\theta_{\rm q}/\beta$. On the right-hand side of \eqref{eq:E12}, only the first term  diverges. 
It is then regularized by the three-dimensional momentum
cutoff $\Lambda$~\cite{Fukushima-1,Fukushima-2,Ratti}.
We use ${\cal U}$ of Ref.~\cite{Rossner}, which is fitted to a LQCD 
simulation in pure gauge theory at finite $T$~\cite{Boyd,Kaczmarek}: 
\begin{align}
&{\cal U} = T^4 \Bigl[-\frac{a(T)}{2} {\Phi}^*\Phi\notag\\
      &~~~~~+ b(T)\ln(1 - 6{\Phi\Phi^*}  + 4(\Phi^3+{\Phi^*}^3)
            - 3(\Phi\Phi^*)^2 )\Bigr] 
            \label{eq:E13}
\end{align}
with 
\begin{align}
a(T)   = a_0 + a_1\Bigl(\frac{T_0}{T}\Bigr)
                 + a_2\Bigl(\frac{T_0}{T}\Bigr)^2,~~~~
b(T)=b_3\Bigl(\frac{T_0}{T}\Bigr)^3  , 
\label{eq:E14}
\end{align}
where the parameters are summarized in Table~\ref{Table1}. 
In pure gauge theory, the Polyakov potential 
yields a first-order deconfinement phase transition at $T=T_0$. 
It is determined from pure gauge LQCD that $T_0=270$ MeV. 
In QCD with two-flavor dynamical quarks at $\mu_{\rm q}=0$, 
the PNJL model with the original value of $T_0$ shows that 
the pseudocritical temperatures 
of chiral and deconfinement crossover transitions are 
$T_{\sigma} \approx 230$~MeV and $T_{\Phi} \approx 215$~MeV, 
respectively~\cite{Ratti}, 
while full LQCD simulations~\cite{KL1994,Karsch4,Kaczmarek2} show 
that $T_{\sigma} \approx T_{\Phi} \approx 173 \pm 8$~MeV. 
It follows from these results that 
the relative difference $\Delta=|T_{\sigma}-T_{\Phi}|/T_{\sigma}$ 
is about $6 \%$ for the PNJL model and at most $10 \%$ for LQCD. 
Thus, for $\Delta$ the PNJL result is consistent with the LQCD data, 
but for the absolute values of $T_{\sigma}$ and $T_{\Phi}$ 
the PNJL result is larger than the LQCD data. 
Therefore, we rescale $T_0$ to 212~MeV in the PNJL model 
to obtain $T_{\Phi}=173$~MeV. However, the PNJL calculation with 
the four-quark interaction only shows that $T_{\sigma}=216$~MeV and 
$\Delta \approx 20 \%$~\cite{Sakai2}. 
The results on $T_{\sigma}$ and $\Delta$ are not consistent 
with the LQCD results. This indicates that the entanglement between the 
chiral and deconfinement transitions is weak in the PNJL model. 
This problem will be discussed later and solved in this paper.

\begin{table}[h]
\begin{center}
\begin{tabular}{llllll}
\hline
~~~~~$a_0$~~~~~&~~~~~$a_1$~~~~~&~~~~~$a_2$~~~~~&~~~~~$b_3$~~~~~
\\
\hline
~~~~3.51 &~~~~-2.47 &~~~~15.2 &~~~~-1.75\\
\hline
\end{tabular}
\caption{
Summary of the parameter set in the Polyakov sector
used in Ref.~\cite{Rossner}. 
All parameters are dimensionless. 
\label{Table1}
}
\end{center}
\end{table}

The variables $X=\Phi$, ${\Phi}^*$ and $\sigma$ 
satisfy the stationary conditions 
\begin{align}
\partial \Omega/\partial X=0. 
\label{eq:SC}
\end{align}
The solutions of the stationary conditions do not necessarily 
yield a global minimum $\Omega$. 
There is a possibility 
that they yield a local minimum or even a maximum. 
We then checked that the solutions yield 
a global minimum when the solutions $X(\theta _{\rm q})$ are inserted 
back into (\ref{eq:E12}). 

The thermodynamic potential $\Omega$ of \eqref{eq:E12} is not 
invariant under the ${\mathbb Z}_3$ transformation, 
\begin{align}
\Phi(\theta _{\rm q}) \to \Phi(\theta _{\rm q}) e^{-i{2\pi k/{3}}} \;,\quad
\Phi(\theta _{\rm q})^{*} \to \Phi(\theta _{\rm q})^{*} e^{i{2\pi k/{3}}} \;, 
\end{align}
although 
${\cal U}$ of (\ref{eq:E13}) is invariant. 
Instead of ${\mathbb Z}_3$ symmetry, however, 
$\Omega$ is invariant under the extended ${\mathbb Z}_3$ transformation~\cite{Sakai}, 
\begin{align}
&e^{\pm i \theta _{\rm q}} \to e^{\pm i \theta _{\rm q}} e^{\pm i{2\pi k\over{3}}},\quad  
\Phi(\theta _{\rm q})  \to \Phi(\theta _{\rm q}) e^{-i{2\pi k\over{3}}}, 
\notag\\
&\Phi(\theta _{\rm q})^{*} \to \Phi(\theta _{\rm q})^{*} e^{i{2\pi k\over{3}}} .
\label{eq:K2}
\end{align}
This is easily understood as follows. 
It is convenient to introduce the modified Polyakov loop 
$\Psi \equiv e^{i\theta _{\rm q}}\Phi$ and 
$\Psi^{*} \equiv e^{-i\theta _{\rm q}}\Phi^{*}$ 
invariant under the transformation (\ref{eq:K2}). 
The extended ${\mathbb Z}_3$ transformation is then 
rewritten as 
\begin{align}
&e^{\pm i \theta _{\rm q}} \to e^{\pm i \theta _{\rm q}} e^{\pm i{2\pi k\over{3}}}, \quad
\Psi(\theta _{\rm q}) \to \Psi(\theta _{\rm q}), \notag\\ 
&\Psi(\theta _{\rm q})^{*} \to \Psi(\theta _{\rm q})^{*} ,
\label{eq:K2'}
\end{align}
and $\Omega$ is rewritten as  
\begin{align}
\Omega = & -2 N_f \int \frac{d^3{\rm p}}{(2\pi)^3}
          \Bigl[ 3 E ({\bf p}) 
          + \frac{1}{\beta}\ln~ [1 + 3\Psi e^{-\beta E({\bf p})}
\notag\\
          &+ 3\Psi^{*}e^{-2\beta E({\bf p})}e^{\beta \mu_{\rm B}}
          + e^{-3\beta E({\bf p})}e^{\beta \mu_{\rm B}}]
\notag\\
          &+ \frac{1}{\beta} 
           \ln~ [1 + 3\Psi^{*} e^{-\beta E({\bf p})}
          + 3\Psi e^{-2\beta E({\bf p})}e^{-\beta\mu_{\rm B}}
\notag\\
          &+ e^{-3\beta E({\bf p})}e^{-\beta\mu_{\rm B}}]
	      \Bigl]+U_{\rm M}+ {\cal U} ,
\label{eq:K3} 
\end{align}
where $\beta\mu_{\rm B}=3 \beta\mu_{\rm q}= 3 i \theta_{\rm q}$. 
Obviously, $\Omega$ is invariant under the extended ${\mathbb Z}_3$ 
transformation \eqref{eq:K2'}, since 
it is a function of only extended ${\mathbb Z}_3$ invariant 
quantities, $e^{3i\theta _{\rm q}}$ and $\tilde{X}(=\Psi, \Psi^{*},\sigma$). 
The explicit $\theta _{\rm q}$ dependence appears only through a factor $e^{3i\theta _{\rm q}}$ in (\ref{eq:K3}). 
Hence, the stationary conditions (\ref{eq:SC}) show that 
$\tilde{X}=\tilde{X}(e^{3i\theta _{\rm q}})$. 
Inserting the solutions back to (\ref{eq:K3}), one can see that 
$\Omega=\Omega(e^{3i\theta _{\rm q}})$. 
Thus, $\tilde{X}$ and $\Omega$ have the 
RW periodicity, 
\begin{align}
\tilde{X}(\theta _{\rm q}+\frac{2\pi k}{3})=\tilde{X}(\theta _{\rm q}), \quad {\rm and} \quad 
\Omega(\theta _{\rm q}+\frac{2\pi k}{3})=\Omega(\theta _{\rm q}), 
\end{align}
while the Polyakov loop $\Phi$ and its Hermitian conjugate $\Phi^*$ have the properties 
\begin{eqnarray}
\Phi (\theta _{\rm q}+\frac{2\pi k}{3})&=&e^{-i2\pi k/3}\Phi (\theta _{\rm q}),
\nonumber\\
\Phi (\theta _{\rm q}+\frac{2\pi k}{3})^*&=&e^{i2\pi k/3}\Phi (\theta _{\rm q})^*. 
\end{eqnarray}

The RW periodicity is a remnant of ${\mathbb Z}_3$ symmetry in the pure 
gauge limit. In QCD with dynamical quarks, 
there appear three ${\mathbb Z}_3$ vacua, 
when $T$ is larger than a critical temperature $T_{\rm E}$. 
The ${\mathbb Z}_3$ vacua are classified by the phase $\phi$ of $\Phi$,  
and each has anyone of $\phi$, $\phi+2\pi/3$ and $\phi+4\pi/3$. 
Roberge and Weiss~\cite{RW} found that there is a first-order phase transition 
at $\theta_{\rm q}=\pi/3$ mod $2\pi/3$ where the ground state 
is changed from a vacuum to its ${\mathbb Z}_3$ images; 
the RW phase transition is illustrated in Fig.~\ref{f6} (shown later). 
The transition is called the "Roberge-Weiss transition." 
In this transition, charge conjugation (C) symmetry is spontaneously broken and $\theta _{\rm q}$-odd quantities such as  the phase $\psi$ of $\Psi$ 
are order parameters 
of the transition~\cite{Kouno}.

In the ordinary PNJL model with the scalar-type four-quark interaction 
only, the chiral transition occurs at higher $T$ 
than the deconfinement transition, 
unlike LQCD data at zero and imaginary $\mu_{\rm q}$. 
In Ref.~\cite{Sakai2}, we revealed that the PNJL model with the scalar-type 
eight-quark interaction~\cite{Osipov,Kashiwa,Sakai}, 
\begin{eqnarray}
G_{\rm s8}[(\bar{q}q)^2+(\bar{q}i\gamma_5\vec{\tau}q)^2]^2, 
\end{eqnarray}
and the vector-type four-quark interaction~\cite{KKKN,Kashiwa,Sakai1}, 
\begin{eqnarray}
G_{\rm v}(\bar{q}\gamma_\nu q)^2, 
\end{eqnarray}
can reproduce LQCD data at imaginary $\mu_{\rm q}$. 
Since the coupling constants, $G_{\rm s8}$ and $G_{\rm v}$, 
of the interactions are adjusted to the LQCD data, 
the correlation between $\sigma$ and $\Phi$ is still weaker in this model 
than in LQCD. 
We then propose another possibility to explain the strong correlation shown 
in LQCD.

\begin{figure}[htbp]
\begin{center}
 \includegraphics[width=0.45\textwidth]{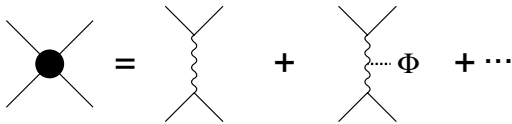}
 \end{center}
\caption{The diagrammatic description of the 
effective vertex $G_{\rm s}(\Phi)$. 
}
\label{f1}
\end{figure}

The origin of the four-quark vertex $G_{\rm s}$ is the one-gluon exchange 
diagram between two quarks and its higher-order diagrams.
If the gluon field $A_{\nu}$ has a vacuum expectation value 
$\langle A_{0} \rangle$ in its time component, 
$A_{\nu}$ is coupled to $\langle A_{0} \rangle$ which is 
related to $\Phi$ through \eqref{eq:E6}~\cite{Kondo}; 
see Fig.~\ref{f1} for the diagrammatic description.
Hence, $G_{\rm s}$ is changed into an effective vertex 
$G_{\rm s}(\Phi)$ that can depend on $\Phi$~\cite{Kondo}. 
The effective vertex $G_{\rm s}(\Phi)$ 
is called the entanglement vertex, and 
all interactions including $G_{\rm s}(\Phi)$ are referred to as 
the entanglement interactions. 
It is expected that the $\Phi$ dependence of $G_{\rm s}(\Phi )$ 
will be determined 
in the future by an exact method such as ERG~\cite{Braun,Kondo,Wetterich}. 
In this paper, however, we simply assume the following $G_{\rm s}(\Phi )$ that 
preserves chiral symmetry, C symmetry~\cite{Dumitru,Kouno} 
and extended $\mathbb{Z}_3$ symmetry~\cite{Sakai}: 
\begin{eqnarray}
G_{\rm s}(\Phi)=G_{\rm s}[1-\alpha_1\Phi\Phi^*-\alpha_2(\Phi^3+\Phi^{*3})]. 
\end{eqnarray}
In the mean field approximation, the mesonic potential $U_{\rm M}$ is 
modified as follows, 
\begin{eqnarray}
U_{\rm M}(\sigma ,\Phi )
=G_{\rm s}[1-\alpha_1\Phi\Phi^*-\alpha_2(\Phi^3+\Phi^{*3})]\sigma^2 
\label{mixU}
\end{eqnarray}
and the constituent quark mass is changed into 
\begin{eqnarray}
M=m_0-2G_{\rm s}(\Phi )\sigma. 
\label{M_modified}
\end{eqnarray}
Thus, this model has entanglement interactions between $\sigma$ and $\Phi$ 
in addition to the covariant derivative in the original PNJL model.
The gap equation (\ref{eq:SC}) can be evaluated by using the chain rules even 
in the presence of the entanglement interactions. 

In this paper, the original PNJL model is simply called PNJL. 
The PNJL model with the entanglement vertex $G_{\rm s}(\Phi)$ is 
referred to as entanglement PNJL (EPNJL), while 
the PNJL model with the scalar-type eight-quark and the vector-type 
four-quark interaction is referred to as PNJL-8V.

\section{Numerical results}
\label{Numerical results}

In this section, we consider the case of $N_{\rm f}=2$, and 
take $m_0=5.5$~MeV unless otherwise mentioned. 
In the PNJL and PNJL-8V models, we take the same parameter set as 
in the previous analysis of Ref.~\cite{Sakai2}. 
In the EPNJL model, we take the same parameter set as the PNJL model, but 
$T_0$ is taken to be 190~MeV so as to reproduce 
LQCD data at $\mu_{\rm q} =0$, when 
$(\alpha_1, \alpha_2)=(0.2,0.2)$. 
This parameter set reproduces LQCD data at zero $\mu_{\rm q}$ 
on the coincidence between 
$T_{\sigma}$ and $T_{\Phi}$~\cite{KL1994,Karsch4,Kaczmarek2} 
and at imaginary $\mu_{\rm q}$ 
on the $m_0$ dependence of the order of the RW endpoint~\cite{D'Elia-3}. 
Qualitative properties such as the coincidence and the $m_0$ dependence 
are preserved in the parameter region 
$\alpha_1, \alpha_2 \approx 0.20 \pm 0.05$. 
The validity of the parameter set in the EPNJL model is confirmed for 
real and imaginary $\mu_{\rm I}$.

\subsection{Transitions at zero and finite quark-number chemical potentials}
\label{quark number}

\begin{figure}[htbp]
\begin{center}
 \includegraphics[width=0.40\textwidth]{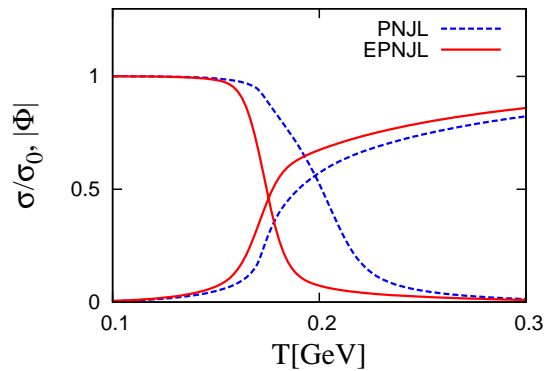}
 \end{center}
\caption{(color online). 
$T$ dependence of the chiral condensate and the Polyakov loop at 
$\theta_{\rm q}=0$. 
The curves that decrease (increase) as $T$ increases represent 
the chiral condensate (Polyakov loop). The solid (dashed) curves are 
the results of the EPNJL (PNJL) model. 
Here, the chiral condensate is normalized by 
the value $\sigma_0$ at $T=0$. 
}
\label{f2}
\end{figure}

\begin{figure}[htbp]
\begin{center}
 \includegraphics[width=0.40\textwidth]{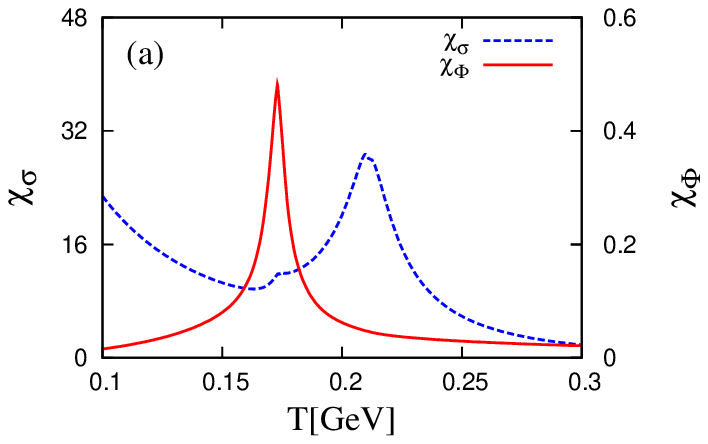}
 \includegraphics[width=0.40\textwidth]{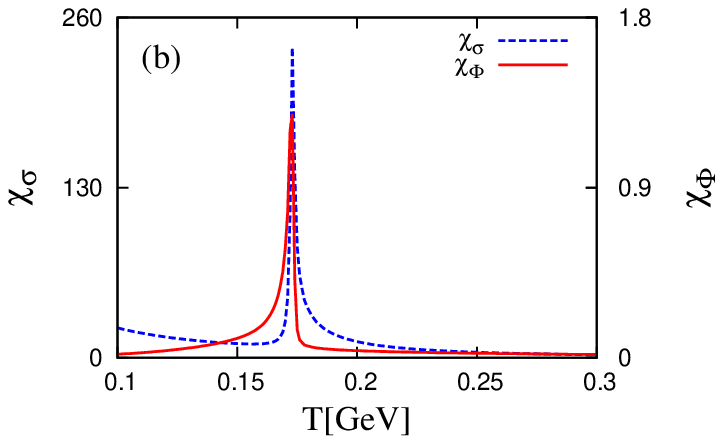}
 \end{center}
\caption{(color online). $T$ dependence of the susceptibilities of 
the chiral condensate (dashed curve) and the Polyakov loop (solid curve) 
at $\theta_{\rm q}=0$. 
Panels (a) and (b) correspond to the PNJL and EPNJL models, respectively. 
}
\label{f3}
\end{figure}

First, we consider the case of $\mu_{\rm q}=0$. 
Figure~\ref{f2} shows the $T$ dependence of the order parameters 
$\sigma$ and $\Phi$, 
while Fig.~\ref{f3} represents chiral and Polyakov-loop 
susceptibilities, $\chi_{\sigma}$ and $\chi_{\Phi}$, as a function of $T$, 
where the susceptibilities are normalized by $T$ to become 
dimensionless~\cite{Fukushima-2, Sasaki-T}.
As shown in Fig.~\ref{f2}, 
the chiral and deconfinement transitions are crossover in both 
the PNJL and EPNJL models. 
Figure~\ref{f3}(a) presents $\chi_{\sigma}$ and $\chi_{\Phi}$ 
in the PNJL model. 
The peak position of $\chi_{\sigma}$, i.e., 
the critical temperature $T_{\sigma}$ of the chiral transition, 
is much larger than the peak position of $\chi_{\Phi}$, that is, 
the critical temperature $T_{\Phi}$ of the deconfinement transition. 
Figure~\ref{f3}(b) corresponds to $\chi_{\sigma}$ and $\chi_{\Phi}$ 
in the EPNJL model. 
In this model, the two transitions coincide with each other within numerical 
errors. 
Thus, the entanglement vertex $G_{\rm s}(\Phi)$ makes the correlation 
between the chiral restoration and the deconfinement transition stronger, 
as expected.

\begin{figure}[htbp]
\begin{center}
 \includegraphics[width=0.40\textwidth]{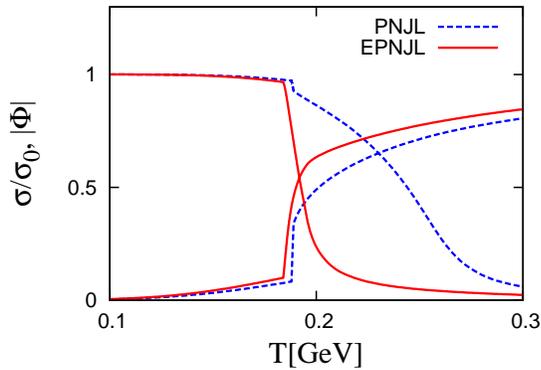}
 \end{center}
\caption{(color online). 
$T$ dependence of the chiral condensate and the Polyakov loop at 
$\theta_{\rm q}=\pi/3$. 
The meaning of the curves is the same as in Fig.~\ref{f2}. 
}
\label{f4}
\end{figure}

Next, we consider the case of $\theta_{\rm q}=\pi /3$. 
Figure~\ref{f4} presents the $T$ dependence of $\sigma$ and the absolute value 
of $\Phi$. In the PNJL model, the deconfinement transition at $T=189$~MeV 
is first order, while the chiral transition is crossover; 
$\sigma$ has a small jump at $T=189$~MeV, but it is just a discontinuity 
induced by the first-order deconfinement transition in $|\Phi|$. 
In the EPNJL model, the deconfinement transition at $T=185$~MeV 
seems to be very weak first-order, 
since $|\Phi|$ has a small jump there 
within the present numerical accuracy, although it is not explicitly seen 
in Fig.~\ref{f4}. 

\begin{figure}[htbp]
\begin{center}
 \includegraphics[width=0.40\textwidth]{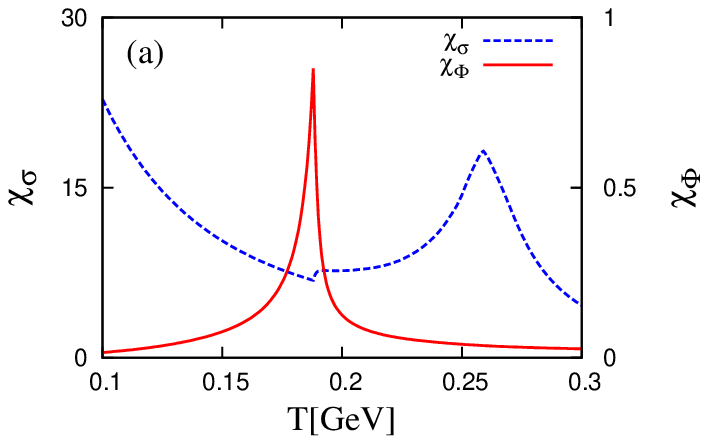}
 \includegraphics[width=0.40\textwidth]{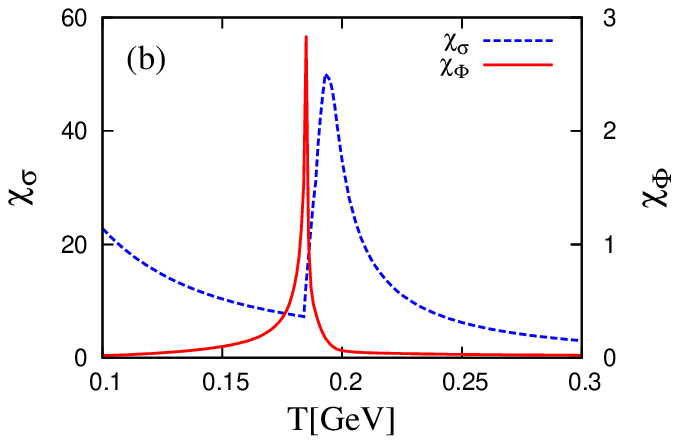}
 \end{center}
\caption{(color online). 
$T$ dependence of the susceptibilities of the chiral condensate 
(dashed curve) and the Polyakov loop (solid curve) at $\theta_{\rm q}=\pi/3$. 
Panels (a) and (b) correspond to the results of the PNJL and EPNJL models, 
respectively. }
\label{f5}
\end{figure}

Figure~\ref{f5}(a) shows that $T_{\sigma} \gg T_{\Phi}$ in the PNJL model, 
while Fig.~\ref{f5}(b) shows that $T_{\sigma} \approx T_{\Phi}$ 
in the EPNJL model. 
Thus, the entanglement vertex yields a stronger correlation between 
the chiral and deconfinement transitions 
also at $\theta_{\rm q}=\pi /3$. 

\begin{figure}[htbp]
\begin{center}
 \includegraphics[width=0.40\textwidth]{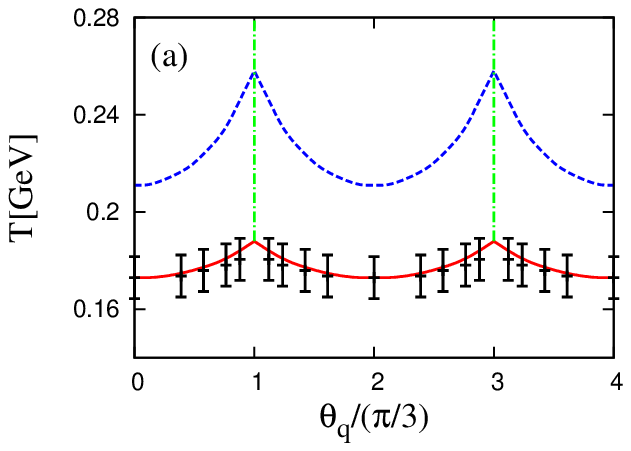}
 \includegraphics[width=0.40\textwidth]{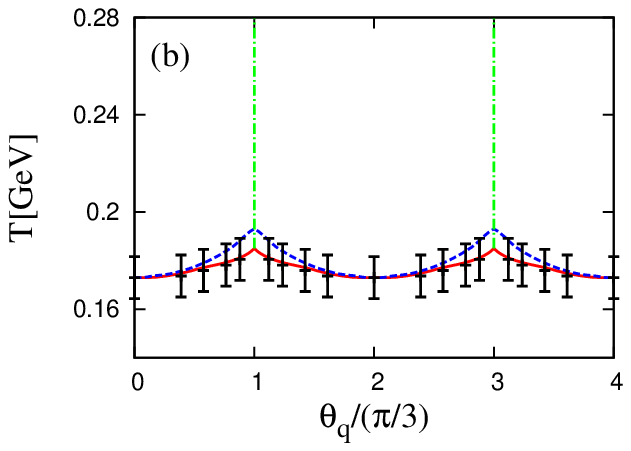}
 \end{center}
\caption{(color online). Phase diagram in $\theta _{\rm q}$-$T$ plane. 
Panel (a) is the result of the standard PNJL model 
with no entanglement vertex, while panel (b) is the result of 
the EPNJL model with $(\alpha_1,\alpha_2)=(0.2, 0.2)$. 
The solid (dashed) curves represent the deconfinement (chiral) transition. 
The vertical dot-dashed lines denote the RW transition lines. 
Lattice data are taken from Ref.~\cite{FP}.
}
\label{f6}
\end{figure}

Figure \ref{f6} shows the phase diagram in the $\theta _{\rm q}$-$T$ plane. 
In the original PNJL model, $T_{\sigma}$ is much higher than $T_{\Phi}$, 
while both are close to each other in the EPNJL model. 
The vertical dot-dashed lines at $\theta _{\rm q}=\pi /3$ mod $2\pi /3$ are 
the RW transition line and the ${\mathbb Z}_3$ images. 
The endpoint of the RW transition line is 
located at $T=T_{\rm E} \approx 189$~MeV in the PNJL model 
and at 185~MeV in the EPNJL model. 
On the RW transition line at $T>T_{\rm E}$, 
C symmetry is spontaneously broken. As a consequence of this fact, 
$\theta _{\rm q}$-odd quantities such as the phase $\psi$ of the modified 
Polyakov loop $\Psi$ are discontinuous, 
while $\theta _{\rm q}$-even quantities have a cusp there
~\cite{Sakai,Sakai1,Sakai2,Kouno}. 
Thus, the $\theta _{\rm q}$-odd quantities are order parameters 
of the RW phase transition. 
In the original PNJL model, the transition is second order~\cite{Kouno} for 
the Polyakov potential proposed by Fukushima~\cite{Fukushima-1}, 
but first order~\cite{Sakai3,Matsumoto} for the Polyakov potential proposed 
by R{\"{o}}{\ss}ner, Ratti and Weise~\cite{Rossner}. 
The latter  is more consistent with LQCD data at imaginary $\mu_{\rm q}$ than 
the former~\cite{Sakai3}. 
In the latter, the deconfinement phase transition is first order near the 
RW endpoint; the endpoint of the first-order deconfinement transition line 
is second order, and susceptibilities of several quantities 
diverge simultaneously there~\cite{Sakai3}. 
In the EPNJL model, such a first-order deconfinement transition line does not 
appear or very short if it does emerge, 
since the deconfinement transition at the RW endpoint 
seems to be a very weak first-order transition, as mentioned above.

\begin{figure}[htbp]
\begin{center} 
 \includegraphics[width=0.40\textwidth]{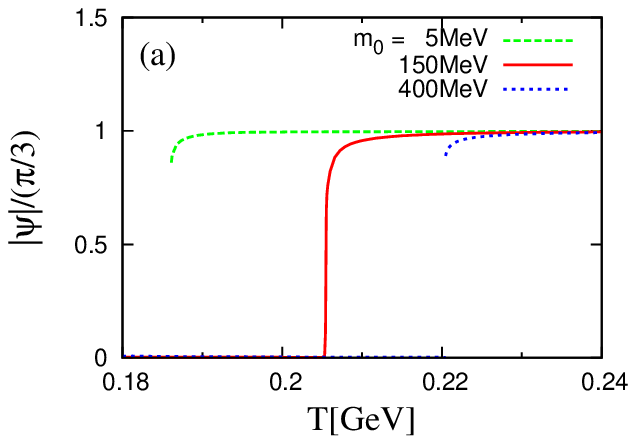}
 \includegraphics[width=0.40\textwidth]{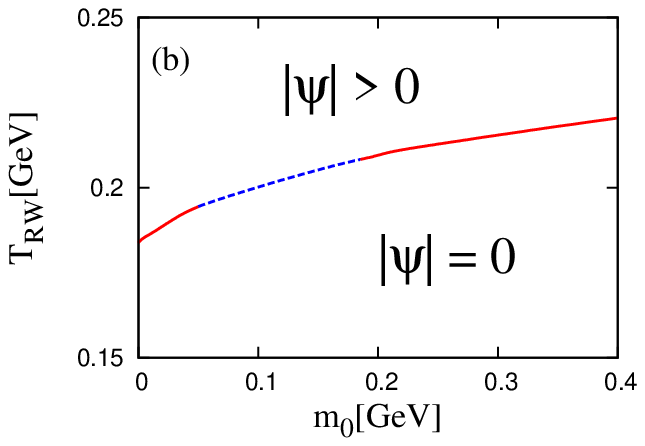}
\end{center}
\caption{(color online). The RW phase transition in the EPNJL model. 
In panel (a), the phase of the modified Polyakov loop 
at $\theta_{\rm q}=\pi/3$ is plotted as a function of $T$ for 
three cases of light, intermediate, and heavy quark masses. 
Panel (b) shows the phase diagram of the RW phase transition in the $m_0$-$T$ 
plane. The solid (dotted) curve shows that the RW phase transition 
on the boundary is first (second) order. 
}
\label{f7}
\end{figure}

Figure~\ref{f7} shows results of the EPNJL model for the RW phase transition. 
Panel (a) presents the $T$ dependence of the phase $\psi$ 
of $\Psi$ at $\theta _{\rm q}=\pi /3$ for the three cases 
$m_0=5,~150$, and 400~MeV. 
The RW transition at the endpoint is first order for 
$m_0=5$ and 400~MeV, but second order for $m_0=150$~MeV. 
In the limit of large $m_0$, the transition is obviously first order, 
since the quark contribution to $\Omega$ is suppressed and hence 
the deconfinement transition is controlled by the Polyakov potential 
$\cal {U}$. 
Meanwhile, the RW endpoint is always first order 
in the original PNJL model~\cite{Sakai3} and in the PNJL-8V model. 
Panel (b) shows the phase diagram 
of the RW phase transition in the $m_0$-$T$ plane; 
C symmetry is spontaneously broken above the curve, while  
it is preserved below the curve. 
The solid (dashed) curve shows that the RW phase transition is first (second) 
order on the boundary. 
The critical mass $m_0(1 \to 2)$ [$m_0(2 \to 1)$]
from the first-order (second-order) to the second-order (first-order) 
transition is rather sensitive to the numerical accuracy. 
In the present numerical accuracy, the critical masses are 
$m_0(1 \to 2)=50 \pm 5$~MeV and $m_0(2 \to 1)=180 \pm 5$~MeV. 
This $m_0$ dependence of the order of the RW endpoint is consistent 
with the recent result~\cite{D'Elia-3,FP2010} 
of LQCD.  

\begin{figure}[htbp]
\begin{center}
 \includegraphics[width=0.40\textwidth]{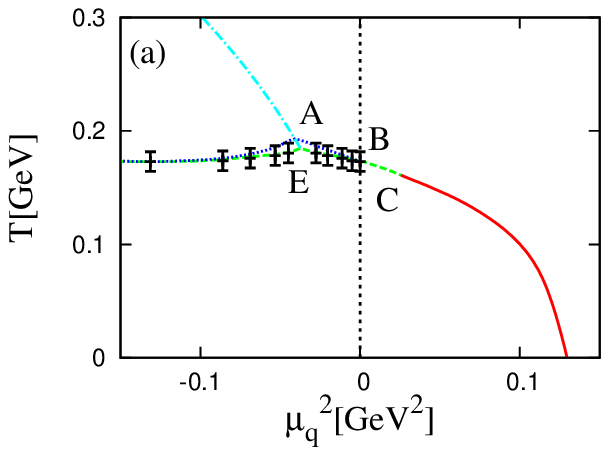}
 \includegraphics[width=0.40\textwidth]{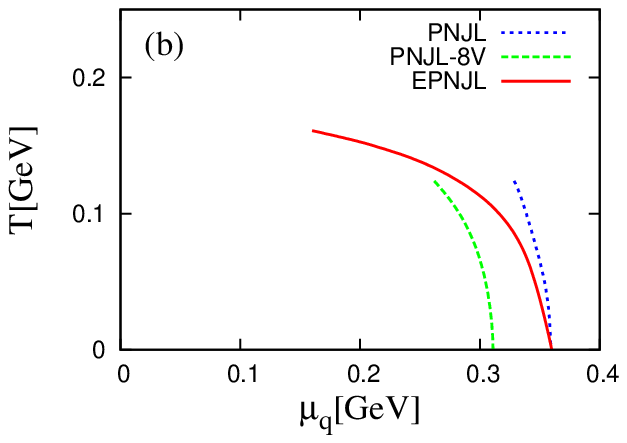}
\end{center}
\caption{(color online). 
(a) Phase diagram in the $\mu^2_{\rm q}$-$T$ plane in the EPNJL model. 
The left (right) half-plane corresponds to imaginary (real) $\mu_{\rm q}$. 
See the text for definitions of lines and points. 
Lattice data are taken from Ref.~\cite{FP}. 
(b) The first-order chiral phase transition line and its CEP in 
the original PNJL, the PNJL-8V, and the EPNJL models. 
}
\label{f8}
\end{figure}

Figure \ref{f8}(a) shows the phase diagram in the whole $\mu ^2_{\rm q}$-$T$ 
plane obtained by the EPNJL model. 
The solid, dotted, dashed and dot-dashed curves represent 
the first-order chiral phase transition, 
the crossover chiral transition, the crossover deconfinement transition 
and the RW transition, respectively. 
The transition line (dashed and solid curves) 
in the region $-0.0375 < \mu_{\rm q}^2 < 0.08$~[GeV$^2$] is expressed as 
\begin{eqnarray}
T=c_0+c_1\mu_{\rm q}^2+c_2\mu_{\rm q}^4, 
\end{eqnarray}
where $c_0=0.173$~[GeV], $c_1=-0.377$~[GeV$^{-1}$], and 
$c_2=-2.71$~[GeV$^{-3}$]. 
Point E is an endpoint of the RW transition, while 
point C is a CEP of the first-order chiral phase transition. 
Point A is a meeting point between the RW transition line and 
the crossover chiral transition line, 
while point B stands for the critical temperature of the 
chiral and deconfinement transitions at zero $\mu_{\rm q}$. 
Locations of these points are tabulated in Table~\ref{Table-ABCD}. 
Thus, there exists a CEP 
not only in the PNJL-8V model~\cite{Sakai2} 
but also in the EPNJL model. 

\begin{table}[h]
\begin{center}
\begin{tabular}{cccc}
\hline
A&B&C&E
\\
\hline
~$(i\pi/3\times 193,~193)$~&~
(~~~0~~~,~173)~&~(160,~161)~&~$(i\pi/3\times 185,~185)$~\\
\hline
\end{tabular}
\caption{Locations $(\mu _{\rm q},T)$ of points A, B, C and E. 
All locations are shown in MeV.
\label{Table-ABCD}
}
\end{center}
\end{table}

Figure \ref{f8}(b) shows the first-order chiral phase transition line and its 
CEP in the original PNJL, the PNJL-8V, and the EPNJL models.  
The locations of the CEP in the three models are summarized in 
Table~\ref{Table2}. 
The CEP is located at smaller $\mu_{\rm q}$ and larger $T$ in the EPNJL model 
compared with the other models. 
Thus, the entanglement vertex yields a drastic effect 
on the phase diagram at real $\mu_{\rm q}$. 

\begin{table}[h]
\begin{center}
\begin{tabular}{llll}
\hline
~~~~~PNJL~~~~~&~~~~~PNJL-8V~~~~~&~~~~~EPNJL~~~~~
\\
\hline
~~~~(327, 124) &~~~~(261, 124) &~~~~(160, 161) \\
\hline
\end{tabular}
\caption{
Summary of locations $(\mu_{\rm q},T)$ of CEP in 
three models. 
All locations are shown in MeV. 
\label{Table2}
}
\end{center}
\end{table}

\subsection{Transitions at finite isospin chemical potential}
\label{Isospin}

The parameter set in the EPNJL model was determined in the previous 
subsection so as to reproduce LQCD data 
at zero and imaginary $\mu_{\rm q}$. 
The validity of the parameter set is confirmed in this subsection for 
real and imaginary $\mu_{\rm I}$ where LQCD data are available.  

The quark-number and isospin chemical potential, 
$\mu_{\rm q}$ and $\mu_{\rm I}$, used in this paper are defined by 
\begin{align}
\mu_{\rm q}=\frac{\mu_{u}+\mu_{d}}{2}=\frac{\mu_{\rm B}}{3},
~~\mu_{\rm I}=\frac{\mu_{u}-\mu_{d}}{2}=\frac{\mu_{\rm iso}}{2}  
\end{align}
with the $u$-quark ($d$-quark) number chemical potential $\mu_{u}$ ($\mu_{d}$).
Here, $\mu_{\rm B}$ and $\mu_{\rm iso}$ are 
the baryon and original isospin chemical potentials 
coupled, respectively, to the baryon charge ${\bar B}$ and to the isospin 
charge ${\bar I_3}$. For comparison with LQCD, we use $\mu_{\rm I}$ as 
the isospin chemical potential instead of the original definition 
$\mu_{\rm iso}$.

The formalism of the PNJL model at finite $\mu_{\rm I}$ is straightforward 
from Sec. \ref{PNJL}. The only essential difference is 
that the pseudoscalar condensate 
$\pi \equiv \langle \bar{q}i\gamma_5 \tau _1 q\rangle$
is nonzero, in general, at finite $\mu_{\rm I}$. 
Therefore, the $E^{\pm}({\bf p})$ in (\ref{eq:E12}) are replaced by 
\begin{eqnarray}
E_{+}^{\pm}({\bf p})=\sqrt{(E({\bf p}) + \mu_{\rm I})^2+N^2}\pm \mu _{\rm q}
\label{up}
\end{eqnarray}
for the $u$ quark and 
\begin{eqnarray}
E^{\pm}_{-}({\bf p})=\sqrt{(E({\bf p}) - \mu_{\rm I})^2+N^2}\pm \mu _{\rm q}
\label{down}
\end{eqnarray}
for the $d$ quark, where $N=-2G_{\rm s}(\Phi )\pi$. 
The mesonic potential $U_{\rm M}$ is also changed into 
\begin{eqnarray}
U_{\rm M}=G_{\rm s}(\Phi) (\sigma^2+\pi^2) .
\label{mesonic}
\end{eqnarray}
See Refs.~\cite{Sakai3} and \cite{Sasaki-T} for the details of the formalism 
with finite $\mu_{\rm I}$; 
the only difference from the formalism is  that $G_{\rm s}$ is replaced by 
$G_{\rm s}(\Phi)$ in the EPNJL model.

\begin{figure}[htbp]
\begin{center} 
 \includegraphics[width=0.40\textwidth]{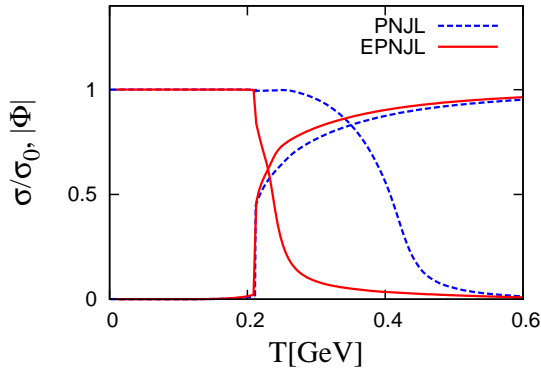}
\end{center}
\caption{(color online). $T$ dependence of the chiral condensate and 
the Polyakov loop at $\theta_{\rm I}=\pi/2$ and $\mu_{\rm q}=0$. 
See Fig.~\ref{f2} for the meaning of lines. 
}
\label{f9}
\end{figure}

First, we consider imaginary $\mu_{\rm I}=i\theta_{\rm I}T$. 
In Fig.~\ref{f9}, we show $T$ dependence of $\sigma$ and $\Phi$ 
at $\theta_{\rm I} =\pi/2$. 
In the standard PNJL model with no entanglement vertex, 
the critical temperature
$T_{\sigma}$ of the crossover chiral transition is about twice 
the critical temperature 
$T_{\Phi}$ of the first-order deconfinement transition.  
This weak entanglement between the chiral restoration and the deconfinement 
transition still persists in the PNJL-8V model also~\cite{Sakai3}. 

The origin of the weak entanglement is the following. 
The $u$-quark loop contribution to the $T$-dependent part of 
$\Omega$ is nearly canceled by the $d$-quark loop contribution. 
Because of this cancellation, the thermal part $\Omega^{\rm th}$ 
of $\Omega$ is reduced 
at $\theta_{\rm I}=\pi/2$ and $\mu_{\rm q}=0$ to 
\begin{eqnarray}
\Omega^{\rm th}&=&-4T\int\frac{d^3{\bf p}}{(2\pi)^3}\left\{{\rm Tr_c}\left[
\ln(1+Le^{-\beta E({\bf p})-\frac{\pi}{2}i})
\right.\right.
\nonumber\\
&&\left.\left.+\ln(1+Le^{-\beta E({\bf p})+\frac{\pi}{2}i}\right]\right\}
+\cal{U}
\notag\\
&=&-4T\int\frac{d^3{\bf p}}{(2\pi)^3}\left\{{\rm Tr_c}\ln(1+L^2e^{-2\beta E({\bf p})})
\right\}+\cal{U}, 
\nonumber\\
\label{iso1}
\end{eqnarray}
because $L=L^{\dagger}$ at $\mu_{\rm q}=0$. 
In the last line of (\ref{iso1}), the exponent in the first term 
is not $\beta E({\bf p})$ but $2 \beta E({\bf p})$, indicating that 
the temperature effect is reduced effectively by 1/2 in the first term. 
In the case of no entanglement vertex, 
the $T$ dependence of $\sigma$ is controlled by the first term, 
while that of $\Phi$ is controlled by $\cal{U}$. 
Therefore, $T_{\sigma} \approx 2T_{\Phi}$ in the original PNJL and the PNJL-8V 
model.

In the EPNJL model, the entanglement vertex appears 
not only in the first term of the last line of (\ref{iso1}) 
but also in the vacuum part of $\Omega$ 
[the first and the $U_{\rm M}$ terms in \eqref{eq:E12}]. 
This induces a strong correlation between 
the chiral restoration and the deconfinement transition. 
Actually, as shown by the solid curves in Fig.~\ref{f9}, 
both the transitions are first order and $T_{\sigma}=T_{\Phi}$. 
LQCD data at $\theta_{\rm I}=\pi/2$ are not available in the 
two-flavor case but in the eight-flavor case~\cite{Cea}. 
The result of the EPNJL model is consistent with the 
LQCD result.

\begin{figure}[htbp]
\begin{center} 
 \includegraphics[width=0.40\textwidth]{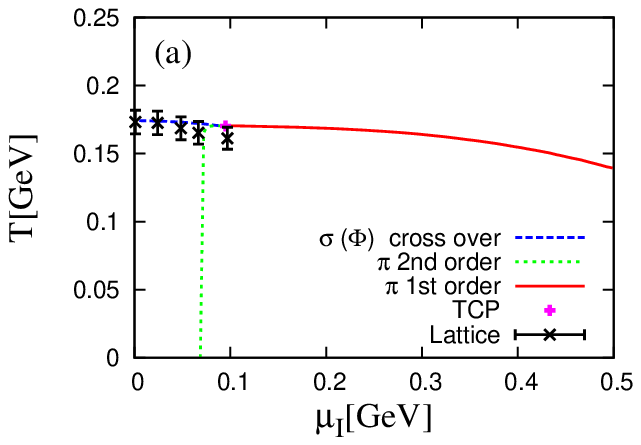}
 \includegraphics[width=0.40\textwidth]{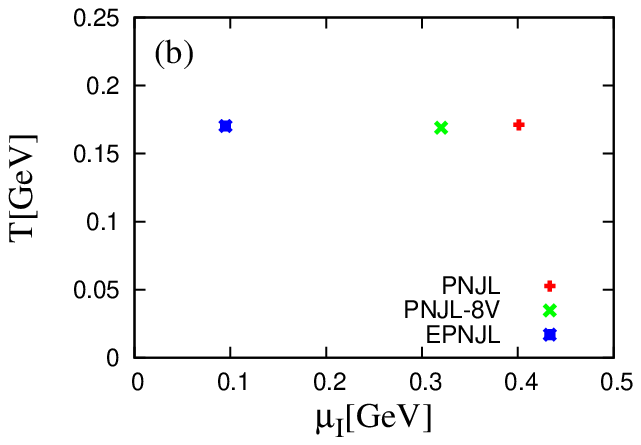}
\end{center}
\caption{(color online). 
(a) Phase diagram in the $\mu_{\rm I}$-$T$ plane at $\mu_{\rm q}=0$ 
in the EPNJL model. 
See the text for definitions of lines. 
LQCD data are taken from Ref.~\cite{Kogut2}. 
(b) The locations of the TCP at $\mu_{\rm q}=0$ in the original 
PNJL, the PNJL-8V, and the EPNJL models. 
}
\label{f10}
\end{figure}

\begin{table}[h]
\begin{center}
\begin{tabular}{llll}
\hline
~~~~~PNJL~~~~~&~~~~~PNJL-8V~~~~~&~~~~~EPNJL~~~~~
\\
\hline
~~~~(401,171) &~~~~(320,169) &~~~~(95,170) \\
\hline
\end{tabular}
\caption{
Summary of locations $(\mu_{\rm I},T)$ of TCP 
at $\mu_{\rm q}=0$ in three models. 
All locations are shown in MeV. 
\label{Table3}
}
\end{center}
\end{table}

Next, we consider real $\mu_{\rm I}$. 
Figure~\ref{f10}(a) shows the phase diagram 
in the $\mu_{\rm I}$-$T$ plane at $\mu_{\rm q}=0$. 
The solid and dotted lines stand for 
the first-order and second-order pion-superfluidity transitions, respectively.
The meeting point between the solid and dotted lines is 
a TCP by definition. 
The crossover chiral and deconfinement transitions agree with each other, 
as shown by the dashed line. 
The EPNJL result reproduces LQCD results 
on the chiral and deconfinement transitions and also on 
the pion-superfluidity transition. 
Figure~\ref{f10}(b) shows locations of the TCP at $\mu_{\rm q}=0$ in 
the original PNJL, the PNJL-8V, and the EPNJL models. 
The entanglement vertex $G_{\rm s}(\Phi)$ largely 
affects the location of the TCP. 
The locations of the TCP in the three models are 
summarized in Table~\ref{Table3}.

\section{Summary}
\label{Summary}

In summary, we have extended the PNJL model by introducing 
an entanglement vertex $G_{\rm s}(\Phi)$ phenomenologically. 
The effective vertex generates entanglement interactions between 
$\sigma$ and $\Phi$. 
The EPNJL model with $G_{\rm s}(\Phi)$
can reproduce two phenomena 
simultaneously; one is the strong correlation between 
the chiral restoration and the deconfinement transition that 
appears in LQCD at imaginary $\mu_{\rm q}$ and real and imaginary 
$\mu_{\rm I}$, and the other is the quark-mass dependence of 
the order of the RW endpoint predicted by LQCD very 
recently~\cite{D'Elia-3,FP2010}. 
Thus, the EPNJL model is consistent with all 
LQCD data at imaginary $\mu_{\rm q}$ and real and imaginary 
$\mu_{\rm I}$.

The functional form of the entanglement vertex 
$G_{\rm s}(\Phi)$ is determined 
by respecting extended ${\mathbb Z}_3$ symmetry, 
chiral symmetry and charge conjugation symmetry. 
The strength of the entanglement vertex is 
determined by LQCD data at imaginary $\mu_{\rm q}$, and 
the validity of this model building is confirmed by LQCD data at 
real and imaginary $\mu_{\rm I}$. 
The entanglement vertex largely changes the location of the TCP 
in the $\mu_{\rm I}$-$T$ plane and the location of the CEP 
in the $\mu_{\rm I}$-$T$ plane.

The present phenomenological approach seems to be 
complementary to the exact renormalization-group approach. 
It is highly expected that the functional form and the strength of 
the entanglement vertex will be determined in the future 
by the theoretical approach.

\noindent
\begin{acknowledgments}
The authors thank P. de Forcrand, A. Nakamura, K. Fukushima, T. Saito and K. Kashiwa for useful discussions. 
H.K. also thanks M. Imachi, H. Yoneyama, H. Aoki and M. Tachibana for useful discussions. 
Y.S. is supported by JSPS Research Fellow.
\end{acknowledgments}


\end{document}